\documentclass[sigconf]{acmart}
\usepackage[section]{placeins} %to place figures exactly where they are intended

\AtBeginDocument{%
  \providecommand\BibTeX{{%
    \normalfont B\kern-0.5em{\scshape i\kern-0.25em b}\kern-0.8em\TeX}}}

\setcopyright{acmcopyright}
\copyrightyear{2025}
\acmYear{2025}
\acmDOI{XXXXXXX.XXXXXXX}

\acmConference[AfriCHI '25]{African conference for Human Computer Interaction}{November 1-6, 2025}{Cairo, Egypt}
\acmBooktitle{AfriCHI '25: African conference for Human Computer Interaction, November 1- 6, 2025}   
\begin{document}
 
\title[Terms and Conditions (Do Not) Apply]{Terms and Conditions (Do Not) Apply: Understanding Exploitation Disparities in Design of Mobile-Based Financial Services}

%%[AUTHORS PLACEHOLDER]
\author{Lindah Kotut}
\email{kotut@uw.edu}
\affiliation{%
  \institution{Information School, University of Washington}
  \city{Seattle}
  \state{Washington}
  \country{USA}
  %\postcode{43017-6221}
}

\renewcommand{\shortauthors}{Lindah Kotut}

\begin{abstract}
    Mobile-based financial services have made it possible for the traditionally unbanked to access infrastructure that have been routinely unattainable. Researchers have explored how these systems have made for safer environments to send and receive money and have expanded financial opportunities such as increased borrowing. With this expansion, challenges such as detrimental interest rates, lack of access to policy documents, and inadequate user protective guardrails emerge, amplifying the risks due to technology-aided unethical financial practices that are aided by design patterns. Supported by user interviews, we detail user experiences of mobile-based financial transactions and explore the foundations and guidelines that undergird the financial service provisions: highlighting both affordances and harms enabled in the design of  such systems. We discuss the findings by highlighting financial exploitation disparities, deliberating strategies for mitigation of risks and enabling recovery from harms caused by the technology use. We then recommend guidelines for empowering design approaches that support users' mechanisms of trust, their understanding of technological processes, and determination of risks.
\end{abstract}

%%
%% http://dl.acm.org/ccs.cfm.
%%


%%
%% Separate the keywords with commas.

\maketitle

%THemes to highlight and the tack to select in addressing the work directions
%TO submit: Critical computing and design theory: Reimagining desing in the face of political/ethical/moral dilemma anddesign research practice. 
%To theme: CRITIQUE/ETHICS/PLURALISTIC/REFLECTIVE/PROGRESS/POLITICS
%To sketch: How MPESA works / Expose the inner functionality and point how the scams and spots that people are taken advatnage of (intro to scam culture)
%Notes: Aproaching this from an ethics perspectives also gives another view of approach for the ethical dimemsion of
%opportunity to better showcase the ethical and historical things that have shifted with MPESA. From only afew people possessing devices (and so ethics of who has access and who control information is a consideration) to now when we seemingly have multiple devices and greater control. There are still others left behind. So we get to talk about it and who is amplified/suppressedin this new reality.

%=============================================
%------- SECTION:  INTRODUCTION --------------
%=============================================

\section{Introduction}
%---------------------------------------------

In envisioning steps to address pressing issues in developing economies (spanning concerns over economy, gender, health and sustainability), and inspired by the new millennium, the United Nations set a 15-year, eight-goal strategy called the Millennium Development Goals (MDG) in 2000 \cite{UNMG}. The first MDG goal was aimed towards halving the rate of poverty  and hunger, and spurring employment \cite{UNMG}. In response to this goal, and to expand access to credit: the M-PESA\footnote{
    M-PESA: ``M'' an acronym for ``Mobile'' and ``Pesa'' a Swahili word for ``Money''} 
platform was developed, piloted, and launched in Kenya. The platform was intended to bridge the financial access gap, and to provide the convenience of mobile-based financial services without the necessity of physical banking infrastructure or a traditional bank account \cite{HughesMIT2007}.

The success of the M-PESA platform was owed to the design approach informed by the understanding of how Kenyans used their mobile phones. For example, the use of airtime as alternative currency to address safety concerns with transporting cash: given that the airtime credit could be transferred across individuals with no fee, and converted to cash without any loss of value \cite{Economist2013}. Since its unveiling, M-PESA has served as a case study for approaching expansion to those without access to financial services (called the unbanked), and users without smartphones, as all the options are available through Short-Message Service (SMS) codes and Unstructured Supplementary Service Data (USSD) menus \cite{SuriScience2016}--making it safer to use than cash for most users \cite{HughesMIT2007}.

With the affordances that M-PESA has provided to marginalized users, new challenges arise resulting from coupling  telecommunication and financial infrastructures. Early research identified user adoption being negatively impacted by the transaction fees and the lack of interoperability across mobile providers \cite{GutierrezWorldBank2014}. The ease of access to finances also led to increased incidences of financial fraud \cite{RazakCSCW2021} and other security-related events \cite{MudiriMicroSave2013}--the people affected by these events often the most vulnerable. Our work builds upon research that have sought different approaches to address these threat concerns spanning research, industry and policy directives, that have identified among others: challenges related to the mismatch between profit-motives and user needs \cite{CsikszentmihalyiCHI2018}, stakeholders lacking the technical know-how \cite{Sultana2021}, and the lack of awareness and easy access to the terms of use and privacy policy documents \cite{MunyendoSP2022}. 

With technological evolution, it has become possible for other services to be built on top of the M-PESA infrastructure, and new mobile loan providers now use the M-PESA platform to lend money and receive payments through the service. While this has led to more people accessing credit, it has also made it possible for the users to be disenfranchised through onerous document and data requirements, and/or be saddled with high interest rates on micro-loans \cite{Hindenburg2020}. The laws and policies intended to protect the users tend to be broad: lacking clarity and scope \cite{FinancialStandard2021}, with enforcement--if mandated, often occurring after demonstrated financial harm/financial loss \cite{TechCrunch2021}. This is both an emerging and ongoing phenomenon, and while enumerated harms are reported in the Kenyan press, the research implications and design guidelines remain broadly unaccounted for in research.

We explored the challenges at these intersections, considering the activities involving the launching of a financial service, the demonstrated abuse, and the effectiveness of responses mandated by policy in addressing the abuse and preventing future re-occurrences. We did so by conducting an interview-based study with 17 participants seeking to understand their experience with mobile-based financial services: against the prescribed guidelines of the UN sustainability goals \cite{UNSG}. We examine user interactions, the trust signals used to mitigate risks--alongside their experience with fraud and financial exploitation, and if/how they recovered from harms. Given this asymmetric context, we use postcolonial computing \cite{IraniCHI2010} to examine the power differentials and colonial traces inherent in design for developmental contexts. In doing so, we examine the types of financial transactions that are conducted, the vulnerabilities faced by the users, and the user awareness on available protective mechanisms--towards seeking design guidelines for supporting trustworthy frameworks that also involve and inform policy. 

We detail the inefficiencies of current tools and systems in protecting and informing the users of existing risks, in mitigating loss of money, and in assigning responsibility for harm repair. We also foreground the demonstrated user resilience (learning, recovering, warning others, etc.) and their custom use of technology. We then discuss the gaps in policy coverage, and opportunities for researchers, designers, and policy makers to address them in support of infrastructures and user knowledge, and to engender accountability and other design considerations in service of underrepresented users in this uniquely African context. 

We make two contributions with this work. First, we highlight the nature of financial-based risk exposure in developmental contexts where there is laxity in policy enforcement and infrastructural support. The resulting user apathy and technology abandonment provide insights to designs intended to support policy, and adds to the nuance of users actions that do not reflect their stated preferences e.g., financial safety. Second, we highlight how users leverage design signals and trust signals to identify risks and evade harms. The findings has implications on HCI researchers who seek to design and envision systems to support users in addressing risks and assist in their harm recovery.

%=============================================
%------- SECTION:  RELATED WORK --------------
%=============================================

\section{Background and Related Work}
%---------------------------------------------

The first of the eight Millennium Development Goals (MDG) presented in 2000 was to ``eradicate extreme poverty and hunger'' \cite{UNMG}--in part by making it possible for people most threatened by poverty to access financial infrastructure. MDG was succeeded in 2016 by the UN Sustainable Development Goals (SDG) whose first goal describe a need to ``promote sustained, inclusive and sustainable economic growth, full and productive employment and decent work for all... [with a target to] ...strengthen the capacity of domestic financial institutions to encourage and expand access to banking, insurance and financial services for all'' \cite{UNSG}. In this work, we follow the path of how this goal has been operationalized in Kenya through the M-PESA platform, the impact on the unbanked, and the opportunities and challenges that have arisen as a result. 

%**********************************************
\subsection{Mobile-Based Financial Services} 
%**********************************************

While the envisioning of the M-PESA platform followed the MDG on reducing poverty by supporting job creation through entrepreneurship, the success of the platform is owed to the confluence of three factors in observance of how the unbanked interacted with money: \textit{transfer and savings, borrowing,} and \textit{alternative currency}. In the absence of banking infrastructure, the dangers inherent in the transfer and saving of cash was understood both by the unbanked--who had limited alternatives--providing the opportunity for M-PESA to supply means of safely doing so. Additionally, the M-PESA platform was chiefly modeled after how people borrowed money through informal and small-scale table banking\footnote{
    Short for ``Cash collected on the table''. This is a funding strategy--often involving women, who meet weekly and contribute a mandated amount (called a share) to a shared kitty. Members can borrow from the kitty at small agreed interest rates. At the end of the year, the interest earned (and fines paid for late fees or late attendance to meetings) are distributed to each member according to their share.} 
systems (called \textit{chamas}) \cite{KariukiCiteseer2014}, and formal microfinance institutions \cite{MutuaFaulu09}. 
%The safety challenges involved in the transfer and savings by the unbanked provided the opportunity for M-PESA to 

The design of the M-PESA platform was also made possible by imitating how the unbanked leveraged cash alternatives for safety, and for access to a quasi-financial infrastructure. The people envisioned ways of overcoming safety and access challenges by re-selling airtime and phone scratch cards, a common practice in developing economies \cite{Economist2013}: as they incur no overhead fees, and can be bartered or used as equivalent currency. All this was aided by the quick adoption of mobile services by Kenyans \cite{MasikaSOUPS2015}; and the eager stakeholder participation in creating new infrastructures in the absence of appropriate policy directives \cite{HughesMIT2007}. 

Researchers have generally supported the M-PESA vision for enabling economic growth and well-being across rural/urban divide, gender, and socio-economic status \cite{DonnerTANDF2008}, and envisioned the need for supporting infrastructures to scaffold the new needs for regulations, and identity management--especially in the face of differing contextual use, differential access to traditional critical infrastructure \cite{KimWiley2018}, and users without conventional forms of proof of identity \cite{MaurerTF2012}. For instance, a retrospective study of M-PESA highlighted factors such as greater access to agents, to financial instruments, and women successfully transitioning from farm work to engaging in commerce in centers \cite{SuriScience2016} to be good metrics of on-the-ground success of the financial platform.

As M-PESA usage expanded, and more was understood about the usage patterns, opportunities arose to compare approaches across different jurisdictions and contexts of use. Researchers have reported on atypical use of devices and technology such as ownership of multiple SIM cards and devices \cite{KotutCHI2024}, and highlighted the lack of platform interoperability \cite{MasamilaIJNSA2014}. More recently, the monopolization of the M-PESA financial infrastructure \cite{CsikszentmihalyiCHI2018} have wrought concerns on fairness, especially to the more financially vulnerable. In this landscape, researchers reported people across different countries were disadvantaged by service fees increased through the use of the platform, and therefore used the mobile payment system to purchase airtime, and little else beside that \cite{GutierrezWorldBank2014, YuCOMPASS2018}--highlighting the practical limitations of the platform in service of the UN SDGs.  As it currently stands, M-PESA is not only a bank, but also a platform, and a service. Its popularity assured through the coupling with Safaricom (the largest mobile provider), and its accessibility across devices from feature phones (leveraging USSD options) to smartphones (through both SIM kits and mobile applications).

With the increased adoption of smartphones, the financial infrastructure access has morphed into the financial technology (fintech) space, where there are new affordances for users to manage their finances \cite{Lewis2019}. The financial applications typically scaffold on the M-PESA platform to enable the transfer of money. This is tightly coupled for users who do not have traditional bank accounts and rely primarily on M-PESA to provide that service.

As the applications in the fintech space have increased in number, researchers have advised caution in their design and distribution: highlighting the harms inherent in the use of western lens to explain non-western contexts that influence behaviors and communal beliefs \cite{Sultana2021}. This has been compounded by the fact that majority of smartphone-based financial applications hold the customer liable for money lost in case of fraud \cite{ReavesTOPS2017}, not accounting for those that occur in transit e.g., man-in-the-middle (MITM) attacks that do not originate from the customer's device. This is underscored by the known difficulty of keeping up with changes to the permission requirements \cite{MunyendoSP2022} and terms of use documents \cite{KotutGROUP2022}: emphasizing the exponential financial loss through fraud--that disproportionately impact people in rural and other marginalized contexts \cite{PervaizCOMPASS2019, IbtasamCOMPASS2018}. 

These reports highlight the scope creep of financial instruments that M-PESA heralded: from a platform to transfer and store money, to a platform that enabled borrowing of money. With these changes, associated challenges and threats have arisen, and form the basis for our approach in using the postcolonial computing lens \cite{IraniCHI2010} to explore how people have interacted with M-PESA as a base platform, related services that are scaffolded on the platform, and other applications in the fintech space in Kenya.

%**********************************************
\subsection{Contextual Implications on Platform Trust} 
%**********************************************

We pull back the lens, and consider how cultural contexts have generally impacted device and technology use. For example, researchers have investigated the impact of communal society and sharing cultures and found nuances involved in the sharing based on needs, gender, communal obligations, etc. \cite{BurrellOUP2010, KotutCHI2024}. African-based studies have found phone ownership to be skewed towards men, the educated, and older members--often from larger households with higher expenditures \cite{BlumenstockICTD2010}. Other findings showcased how sharing patterns varied across friends, family groups \cite{MurphyTF2011} and with generational variance \cite{KotutWoC2022}. Device usage was also found to be impacted by access to traditional infrastructure (e.g., connection to the electrical grid), and the cost of maintenance of devices \cite{WycheDIS2012}. In the presence of these structural barriers, phone ownership also tended to amplify the existing inequalities--more than it served to address poverty \cite{MurphyTF2011, Kotut2020Griot}.

The notion of personal privacy in the context of a sharing culture has also served to reveal the nuances of gender--given geographical restrictions on movement across socio-economic and education status \cite{IbtasamCOMPASS2018}. This has foregrounded the need to support contextual privacy/data secrecy boundaries, and to account for the shared use paradigm \cite{AhmedCSCW2017, KotutCHI2024}--with implications on the customer policy data requirements \cite{GitauCHI2010, KotutGROUP2022}, and government-mandated biometric registrations, raising concerns on surveillance \cite{AhmedCHI2017}. 

Given the challenges of identity management in the face of shared device culture, researchers have considered how people establish trust--especially in spaces where there are often ethically dubious practices and lax protection of personal information \cite{BowersWISEC2019, ReavesTOPS2017}. Studies have explored user-inferred flags on smartphone applications: popularity, customer rating, presence of privacy policy, etc., as a predictor of over-privileging of applications \cite{ChiaWWW2012}. The findings reveal nuances, that while there was no one reliable predictor of trust, the key enablers of mobile fraud tended to be influenced by a pipeline of events: lack of informed consent, lack of education, and a sense of shame in reporting fraud \cite{PervaizCOMPASS2019, RazakCSCW2021}. This is further compounded by poor communications, weak standards, and lax regulations--which in turn impact all stakeholder relationships, from the customer, to the agent, to the mobile provider \cite{MudiriMicroSave2013, NamaraAfriCHI2018}. Trust signals, when combined with financial literacy and practices, highlight the need to design for trust:  for instance through validating successful reversible/irreversible transactions and providing transparent fee structures \cite{BailurInteractions2021}.

We center these insights in this work--examining user actions in the face of these policies, seeking to understand the structural concerns: how users leverage their contextual trust signals given the protections they require; and how we might approach design to account for these factors. We also seek to understand where these trust signals fail, how they have evolved, and in case of acknowledged harm through fraud, when, how, or if trust was recovered. 

%---------- Figure: Timeline --------------
\FloatBarrier %to force the figure placement
\begin{figure*}[!ht]
    \centering
    \includegraphics[width=1\textwidth]{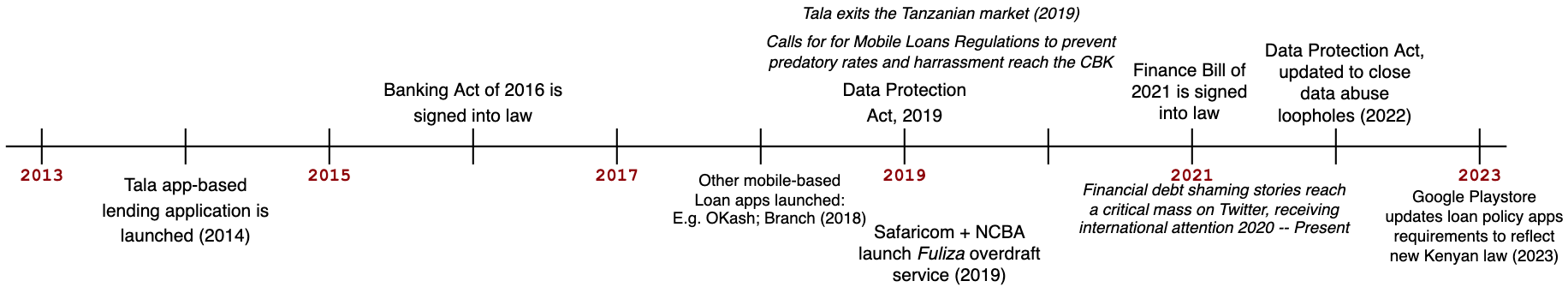}
    \caption{A ten-year timeline that highlights the gaps between the initial offering of a financial-based mobile application, when loopholes and harms were evidenced, and when policies/laws were updated or enforced.}
    %\Description: A 2013-2023 timeline of the evolution of mobile-based loan providers in Kenya. Timenline offers a highlight of when key players entered the market, and when specific policies were enacted. The figure represent a summary of section 4.1.
    \label{fig:timeline}
\end{figure*}
%---------- /Figure: Timeline --------------

%**********************************************
\subsection{Privacy and Security Considerations}
%**********************************************

Given the focus on trust and finances, and the sensitivity of information collected from users, we are also informed by research on privacy and cybersecurity scoped within the HCI for development umbrella. Researchers have considered this intersection of users, platform/app security hygiene and adversarial impact, to understand threat ramifications and the gaps they reveal, including user awareness and education and lax/unrevised policies \cite{Ben-DavidNSDR2011}. Instrumentation studies have also highlighted the variability in the security and privacy standards, leading to the prevalence of unsecured mobile-based financial applications \cite{AhmedAccess2021, ChenICSE2020, BowersWISEC2019}. Other approaches have considered  the problem from the adversarial incentives angle, seeking to understand how people fall for financial scams \cite{StajanoACM2011}, in an effort to inform the design of resilient systems. 

The lack of formal policy and privacy frameworks to inform interoperability \cite{MasamilaIJNSA2014}, and financial regulation requirements (e.g., ``know-your-customer'' (KYC) compliance) to protect collected personal information, may expose customers to targeted advertising and abuse \cite{HarrisHein2012}. Cultural practices, knowledge gaps, unintended technology use, and long term use of devices have also added complexity and challenged the traditional security/privacy models \cite{VashisthaCOMPASS2018}. 

Users are burdened with having to find responsive channels to report noncompliance \cite{ShahICTD2019}, in addition to navigating the dearth of educative sources to inform how they defend themselves from fraud \cite{JainCOMPASS2021}. Application developers are burdened with devising structures to inform how they develop technology and handle user information in the financial space: often resulting in patchwork  placeholder frameworks \cite{HughesMIT2007} that are difficult to replicate, and may be prone to obsolescence \cite{CastleDEV2016} and abuse \cite{BowersWISEC2019}. These burdens make it more likely that the mobile-based financial services use outdated security practices, lack coherence between disclosures provided in the policy document(s) and the actual actions conducted, and lead to abdication of responsibility in cases of harm. 

We focus on the harms in the contextual description below, with the aim of offering a snapshot of the current landscape given the policy historical imperatives. We expound on how adversarial methods have evolved given the changes in policy, how the users have been affected as a result, and the mechanisms by which they use to recover from harm--with implications on design, policy, and research.

%=============================================
%-------- SECTION: CASE STUDIES --------------
%=============================================

%\section{Case Studies}
%---------------------------------------------

%We present two case studies on the financial affordances available through borrowing opportunities in the technology-enabled financial landscape in Kenya, and emphasize (in bold type) the burdens the users face. We then present example contexts where users encountered financial threats: spanning traditional financial-based fraud to targeted financially-grounded deception practices. Both case studies are intended to provide insights to discuss institutional and personal impact: touching on aspects of law, policies, fairness, justice, privacy, security and trust. This approach has implications on designing technology for use, and the responsibility to be borne by each stakeholder in the design. 

%**********************************************
\subsection{Context: Mobile-Financial Landscape in Kenya} %RQ1, RQ2, RQ3 %INSTITUTIONAL ABUSE and the new underclass
%**********************************************

Micro loans are in the spirit of UN SDGs--enabling marginalized groups and others without credit histories to have access to loans at reasonable interest rates: avenues for lifting themselves out of poverty \cite{UNSG,HughesMIT2007}. This started in the early 90's in Kenya with micro-finance institutions such as the Faulu Microfinance Bank and the Kenya Women Microfinance Bank (KWFT) both offering micro loans in addition to financial awareness training \cite{MutuaFaulu09}. The institutions were formalized within the law in 2006 \cite{CBK-Microfinance-Act-2006}, and thereafter, the ubiquity of mobile phones and subscription to the M-PESA service made it possible for the microfinance infrastructure to be offered through mobile devices. These mobile-based unsecuritized\footnote{
    Unsecuritized means that there are no traditional forms of collateral used to secure the loan, making it a higher risk for the entity offering the loans. The risk is offset by the lower amount offered and higher interest rates charged.} 
loans are typically between Ksh. 100 and Ksh. 20,000\footnote{
    Roughly  \$1 to \$200 in US Dollars}, 
with a 30-day payment period at roughly 7\% monthly compounded interest rate (when annualized, can translate to roughly 90\% interest--compared with $\sim$13\% per annum that is the threshold set by the Central Bank of Kenya for regular bank loans) \cite{CBK}.

In 2012, M-Shwari--a USSD-based loan service was released by the mobile provider Safaricom (serviced through Commercial Bank of Africa) as an online-only service \cite{M-Shwari}. This provided an opportunity for subscribers who used M-PESA for storing and sending money, to become eligible to borrow loans. The leveraging of Safaricom's wide reach across the country and the existing infrastructure, as well as the users' familiarity with the M-PESA interface, made the program a success. This led to other USSD and venture-backed app-based competition: Tala in 2014 \cite{Tala}, followed in 2018 by Branch, OKash and others \cite{Hindenburg2020}. Notably, \textbf{there was lack of sufficient oversight for these new financial  offerings}, and the market was rife with prohibitively high interest rates and short repayment periods that disadvantaged the very same people who stood to most benefit from access to microloans.

As a remedy for the unfair interest rates and the disproportionate impact this had on marginalized groups, the Kenyan government enacted the Banking (Amendment) Bill of 2016 \cite{CBK-Banking-Act-2016}: limiting the interest rates to not exceed the threshold set by the Central bank of Kenya (CBK). However, this failed after a court challenge by the financial institutions--who argued that mobile loans were exempt from the interest caps set by the new law, since they were classified as microfinance entities: their terms agreement showcasing that they charged ``facilitation fees'' and not the traditionally understood ``interest rates''. In practice, this also meant that unpaid micro loan balances  were also charged a ``roll-over fee'' (essentially a compound interest) after 30 days \cite{BusinessDaily2016}--at a rate that was far above what was set by the CBK. This was done with the knowledge that \textbf{feature phone users had no access to these online policy documents}, and further led to onerous burdens on people who often found themselves in debt almost twice the size of the original loan.

Users who sought app-based microloans were usually required to surrender their contact list and transaction history as conditions for loan approval. \textbf{Access to these details were often abused by the lenders}: as they called the people in the borrower's contact list to ask them to relay messages to the borrower to repay their defaulted loan. This practice of shaming borrowers and the general terrible user experiences created enough national and international furor \cite{TechCrunch2021, MaringaDW2023, MwauraBBC2021, OdhiamboEA2022, RoussiFT2020} that the Banking Act was amended in 2021, and the CBK was formally placed by law as the governing authority over all loan-provision services including mobile-based microfinance loans \cite{CBK-Banking-Act-2021}. This finally closed the five-year loophole on the microfinance exemptions on interest rates. In addition, the Data Protection Act  was amended to curb the abuse of personal and financial information \cite{ODPC-2021}, and additionally required the digital lenders to submit a proof of license to operate in Kenya--extending the CBK's governance to include loan-service apps that are available in the app stores \cite{NjanjaTC2023}. Figure \ref{fig:timeline} provides a timeline overview.

\begin{table*}[hbt]
\small{
\begin{tabular}{llrl||llrll}
 \hline
 \textbf{Code}  & \textbf{Gender} & \textbf{Age} & \textbf{Occupation} & \textbf{Code}  & \textbf{Gender} & \textbf{Age} & \textbf{Occupation}\\
 \hline
    P1  & Woman (W) & 18 & Student &                     P11  & M & 63 & Retired Office Worker         \\
    P2  & W & 20 & College Student &             P12  & M & 68 & Business Man                  \\
    P3  & Man (M) & 24 & University Graduate &         P13  & W & 72 & Retired Civil Servant         \\
    P4  & M & 25 & Shopkeeper &                  P14  & W & 79 & Volunteer (Wildlife Services) \\
    P5  & M & 28 & Entrepreneur (Livestock) &    P15  & W & 81 & Entrepreneur (Hospitality)    \\
    P6  & W & 32 & Museum Docent &               P16  & M & 86 & Farmer                        \\
    P7  & M & 41 & (Religious) Ministry Worker & P17  & M & 91 & Retired Farmer                \\
    P8  & M & 53 & Driver                                                                      \\
    P9  & W & 60 & Green Grocer                                                                \\
    P10 & W & 61 & Retired Teacher                                                             \\
    \hline \hline
\end{tabular}}
\caption{Summary of interview participants ranked by age (from 18 years to 91 years). The occupations are self-described by participants. We include addendum in parenthesis to provide additional context.}
\label{tab:participants}
\end{table*}
%------------- /TABLE --------------

%=============================================
%-------- SECTION: METHODOLOGY  --------------
%=============================================

\section{Method}
%---------------------------------------------

We interviewed 17 participants (8 women, 9 men), aged between 18 and 91 years (Median 53.1) across three locations in Kenya: Nairobi (the capital city), Mombasa (a coastal city) and the rural areas surrounding the highland town of Eldoret. The participants were recruited through convenience sampling--from quick surveys and word of mouth--the first author having had a long-term research relationship in various locations in Kenya. We sought participants who had had experiences with, or had been witness to attempted and/or successful tech-enabled harms, and we further restricted this to those who had borrowed money using mobile-loan services and were aware if not on exact interest rate, then the exact amount of money either deducted from the initial loan remitted or the extra amount required to be repaid. 

The interviews provided opportunities to incorporate participants in urban and rural contexts; those who had different familial responsibilities: with variance in literacy skills, socio-economic status, and technological know-how. The interviews were conducted either in person at locations we set up, or in public settings according to participants preferences e.g., shopping centers, cafes, or a participant's place of business (Table \ref{tab:participants} provides a summary). When feasible, participants were encouraged to share screenshots to support their interview--these were collected through a survey \cite{DoganMPESA2025}. The study underwent institutional ethics review and approval.

We sought to understand how the participants leveraged mobile-based financial transactions using M-PESA and other applications involving money transfer, savings and loans. We were interested in their understanding of the nuances of interest rates and penalties for late repayment, in addition to other experiences of lost money (or near misses) due to outright fraud. We iteratively sought responses with each participant until we could no longer add new experiences and/or noted multiple redundancies in their responses. In elucidating their experiences, the participants offered insights on how they coped with threats/known harms, and if applicable and available, the tools and resources they leaned on for support and guidance. 

Each interview lasted between 20 and 50 minutes, and was conducted in English or Kiswahili according to a participant's preference, and was also audio recorded with permission. The interviews were led by the first author who is Kenyan born and raised: with fluency in both languages and their variance--the background serving to engender trust, and support open conversations with the participants. Each participant received Ksh. 500 ($\approx$\$3 USD) in remuneration.

%**********************************************
\subsection{Analysis}
%**********************************************

We assembled a written corpus to use in the analysis. The corpus consisted the transcribed and translated transcripts from the audio files, additional notes from the interviews and the descriptions of participant-contributed artifacts (e.g., SMS Screenshots). We used open coding to elicit concepts and subsequently grouped them into categories that help to explain themes based on participants' experiences \cite{CorbinStrauss2023}. We then supported the themes using contextual details given the mobile-based financial landscape in Kenya, as detailed in Section 2: framing the participants' experience given the background and insights surrounding the Kenyan financial landscape that is otherwise inappreciable when considering individual/isolated situations. We discuss our findings using postcolonial computing \cite{IraniCHI2010} as lens. This approach provides the space to contend with colonial influence on the designs, contrast cultural approaches to technology use, and the power asymmetries inherent in the technology design and infrastructure enabled by technology.

%============================================
%---------- SECTION: FINDINGS I -------------
%============================================

\section{Findings} %Impact on Trust and Culture
%---------------------------------------------

Through their awareness and experience with the M-PESA infrastructure, the participants articulated their means of determining trust signals of legitimate transactions (\textit{origin signals, spam filter}s and \textit{claim verification}), and how they navigated and/or understood how others navigated through similar issues--including the impact to personal dignity if tricked by deceptive transactions.

%First we explore the repercussions of financial stature given the burdens highlighted in Case I: High interest rates, lack of access and understanding of policies, and the abuse from loan providers--towards understanding how participants coped. %and the implications on behavior, agency and privacy.

%**********************************************
\subsection{Awareness and Verification}
%**********************************************

\begin{figure*}  %[ht]
 %\FloatBarrier %to force the figure placement
    \centering
    \includegraphics[width=0.80\textwidth]{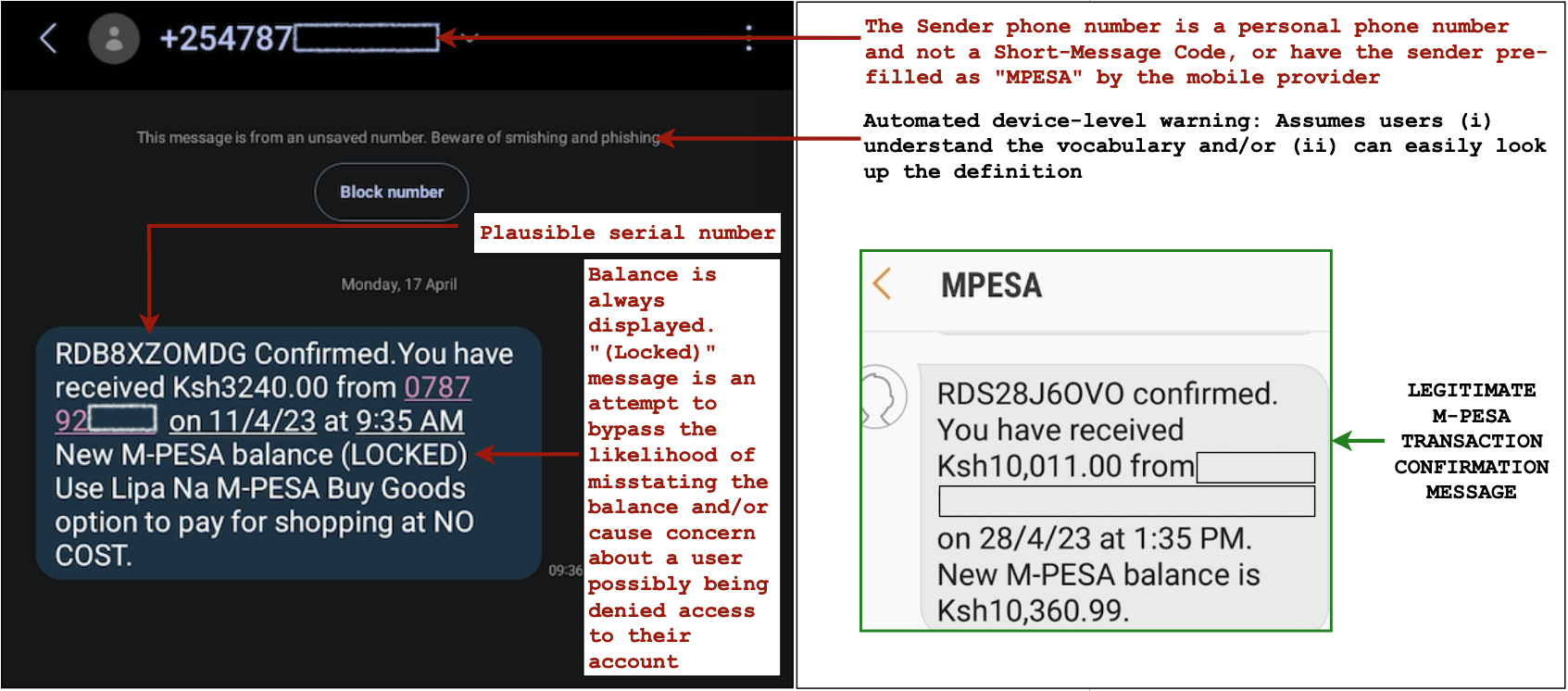}
    \caption{Example of an M-PESA scam: Fake confirmation deposit originating from a personal number. An inset text message showcases what the transaction details provided in legitimate confirmation message. }
    \label{fig:mpesa-scam}
    %\Description: Figure shows a screenshot of SMS that reads "RDB8XZOMDG Confirmed. You have received Ksh3240.00 from 078792[...] on 11/4/23 at 9:35 AM New M-PESA balance (LOCKED) Use Lipa na M-PESA Buy Goods option to pay for shopping at NO COST". Added commentary to the figure shows that the serial number is plausible, and the "Locked" message takes advantage of those who do not understand how typical M-PESA balance is displayed.
\end{figure*}

%he general patterns surrounded: \textit{Origin signals, device/platform-provided signals} and \textit{claim verification}.
Participants utilized various signals to distinguish between legitimate and deceptive financial-related messages--with most implementing different aspects of \textbf{origin signals}. For Safaricom subscribers, the short code assigned to M-PESA system often has the name ``MPESA'' as pre-filled contact without having to be saved in a subscribers address book (as shown in Figure \ref{fig:mpesa-scam}).

\begin{quote}
    \textit{``I didn't know at first glance it was a scam until I realized that it was a message from an individual number, not the official M-PESA number...'' (P14) ``[...] it actually seemed real but the message itself was not sent through the M-PESA code, and then instead of showing you the balance in the message, it instead said `locked': definitely a scam.'' (P1)}
\end{quote}
The M-PESA short-code message service supports only one-way (outgoing) communication. Some participants classified messages as deceptive if they did not follow expected behavior of legitimate messages. 

\begin{quote}
    \textit{``I noticed two things: the message came from a personal phone number, and also supported replies.'' (P6)} 
\end{quote}
Other participants leveraged \textbf{device or platform-provided} spam filters supported natively in their devices or through third-party applications, to automatically identify deceptive messages:

\begin{quote}
    \textit{``My phone has an internal spam detector so the message was automatically archived as spam ...'' (P5) ``[...] I have Truecaller\footnote{A third-party crowd-sourced caller ID service. \url{https://www.truecaller.com/}} installed on my phone and some of these messages come directly into my spam folder in Truecaller.'' (P2)} 
\end{quote}
Participant approaches that leveraged \textbf{claim verification} did so in two ways: first by verifying their balance directly through typical practice, for example: \textit{``The message said I had received money in my M-PESA account but when I checked my account there wasn't any money'' (P7)}. Alternative approaches verified claims by employing third-party verification: 

\begin{quote}
    \textit{``The message indicated that I should deposit my rent to a new account number and that it was a change from the management. I had just spoken to them in person in the past week and they hadn't said anything so when I called to understand the change, that is when I found out that they were not the ones who had sent the text, and it was probably a scam. The message was very believable though.'' (P8)} 
\end{quote}
The experience by P8 evidences both the changing deceptive strategies and the limited knowledge-base that users have to determine veracity. For example, while the participants could easily identify the parent/child deception category (\textit{``The sender wanted me to send money to buy a calculator for a son at a certain high school, but I'm not a parent'' (P2)}), whenever the deceptive message better matched a user's context, determination was murky.

\begin{figure*} [ht]
 %\FloatBarrier %to force the figure placement
    \centering
    \includegraphics[width=0.80\textwidth]{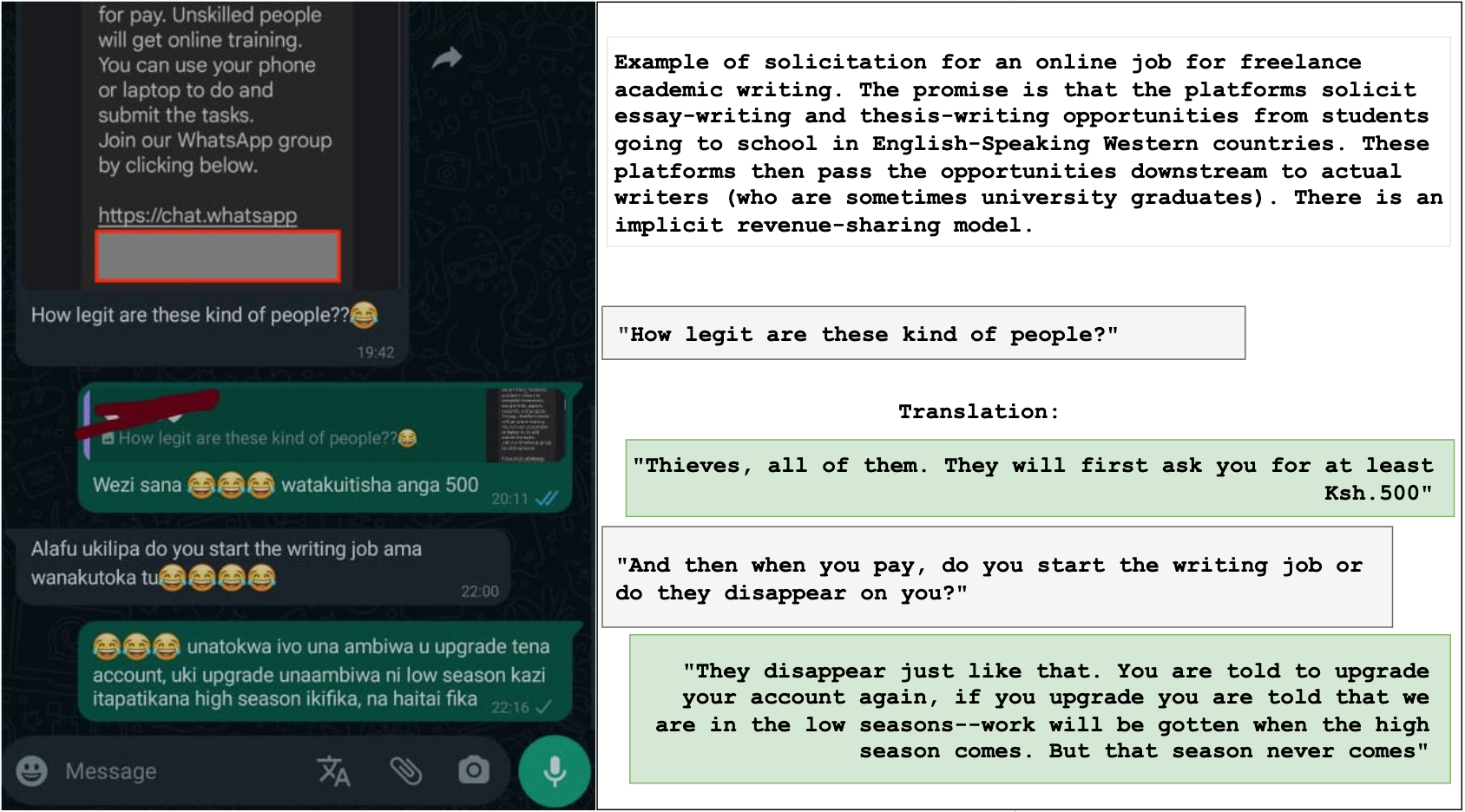}
    \caption{Participant-shared example of how a young adult leveraged their trusted circle to determine the trustworthiness of a job opportunity in an ethically-gray area. They sought to understand the deceptive approaches and potential financial loss.}
    %\Description: Figure 4: Screenshot of an M-PESA account showcasing conversation with two people discussing an Academic writing job opportunity (an ethically-dubious scheme where a service is offered to student in western-contexts who need essay-writing support). Someone asks about the legitimacy, and the responder explains how they ask for Ksh 500 as an upfront fee, but would instead ask for more money to upgrade, then finally telling the user to try again during the high season, as they are suffering from low number of available opportunity. 
    \label{fig:academic-writing-scam}
\end{figure*}

\begin{quote}
    \textit{``They had texted me that the rent paying account number had changed. I almost fell for it because I didn't have the number of our caretaker so I tried calling the number to confirm, but it was unreachable.'' (P4)}
\end{quote}
%
%---------- Figure: Scam --------------
%---------- /Figure: Scam --------------
%
%The classification on personal touch expanded from verification steps undertaken by participants. Examples included deceptive attempts pretending to ask for help from a relative:
%
%\begin{quote}
%     \textit{``Claimed to be my relative. I asked them to verify their nickname, and they couldn't. A scam.'' (P12)}
% \end{quote}
%
Other deceptive approaches take advantage of the likelihood of young adults to be job searching and expected involvement of financial corruption to craft believable messages. While there are circumstances where verification using trusted circles (as exemplified in Figure \ref{fig:academic-writing-scam}) is possible, this is not uniform.

\begin{quote}
    \textit{``I went through an interview process and was very hopeful that I would get the job, but then they texted me after the call with the same number requesting for money for me to get the job. I thought it was just the usual corruption stuff, but I found out later that this is a common scam, and they didn't actually have any jobs available.'' (P3)}
\end{quote}
Deceptive practices extended to approaches accounting for other demographic characteristics and their likely financial-related practices such as M-PESA supported betting: a common practice aided in part by majority of radio and TV stations that offer betting-related trivia competitions.

\begin{quote}
    \textit{``They actually called me and said they were from [radio station] and that I had won the betting sweepstakes, but that I should first send some money to a paybill number before they would then send the rest of the money. It is true that I participated in the sweepstakes, but who would ask you to pay money to then get more money? It didn't make any sense.'' (P9)}
\end{quote}
The various means of verification were based on user-elicited signals. While the participants were aware of mobile-providers campaigns to create awareness for deceptive practices, the specifications and strategies provided in these campaigns did not feature in the participants' own practices. 
%``The message came from an ordinary number requesting it's a service number until I had to contact Safaricom about it''

%**********************************************
\subsection{Responsibility, Dignity, and Shame}
%**********************************************

Participants also employed different strategies for navigating mobile-enabled micro loans. Emergent themes surrounded \textbf{dignity} and \textbf{shame}--chiefly in the context of borrowing loans. By virtue of surrendering access to their contact list as a condition for mobile loan approval, borrowers were liable to experience any person on the list being called by the loan provider as a tactic to compel and shame the borrower into repaying the loan. Shame as a theme also extended to participants' feelings after having been defrauded. 

Older participants were also likely to have abbreviated formal education and experience with the M-PESA USSD menu. One participant (P16), beyond noting that they had lost money, declined to articulate how they were defrauded and the amount they lost. 

\begin{quote}
    \textit{``I can't tell you. It was too bad [...] it led me to lose a lot. My family does not know this, actually I've never told anyone. I don't want them to know about it.'' (P16)} 
\end{quote}
The success of deceptive strategies that focused on micro-loan provision can be attributed to those with no borrowing history--often first time users of loan offerings through M-PESA. These users have little awareness over fair rates and had no choice but to accept onerous repayment requirements. They are also more likely to hesitate reporting their loss, as they often deem their cases as too insignificant to be taken seriously by the Directorate of Criminal Investigations (DCI)--the government agency with oversight over fraud investigations.

\begin{quote}
    \textit{``They sent me a message claiming that they were giving away [Bank Name] loans at an interest of 20\% and they give you a number to call. When I called, they told me to send money as a service fee to them for the loan to be disbursed. It was a reasonable fee, especially given the loan I was to receive [...] But then I never heard from them again.'' (P1)}
\end{quote}
Neither the farmer (P16) nor the student (P1) reported the numbers used to deceive them to authorities. The student knew the means to report to the service provider, but felt that since the mobile provider did not have the reach and effectiveness of the government agency, they would not be of help in restoring the money lost.

%**********************************************
%\subsection{Behavioral implications}
%**********************************************

With the low reporting incidences and assumptions to navigate unassisted through their losses, the participants compensated for their lack of control by withdrawing to known maxims: elevating their technology distrust levels and/or reverting to proven verification means. For some, harassment through unceasing phone calls led them to revert to solely relying on locally-organized microfinance structures, abandoning mobile-based services altogether. 

\begin{quote}
    \textit{``I got tired of the phone calls [from mobile-based loan providers]. Now I would rather borrow from my ``chama'' [table-banking group] than to go through all of that again.'' (P11)}
\end{quote}
Other participants became wary of unsaved numbers as they had previously fallen victim to the financial fraud. They considered the scammers as untouchable and had no recourse in stopping the harassment. 

\begin{quote}
    
    \textit{``The scammers claim to be the true agents because they have your information. But they use vulgar language when you catch on to their trick. I don't think they care if you report them.'' (P10)}
\end{quote} %``It wasnt a safaricom number and the sender became aggresive once I inquired how they required my details to get back the money yet they could just reverse it''

%---------- Figure: Academic Writing --------------
%---------- /Figure: Academic Writing --------------

%
%**********************************************
%\subsection{Technology and Policy Implications}
%**********************************************
The relative age of the two participants (P10 and P11) above also reflect the general patterns of behavior we observed across the rest of the participants. The older participants often had one phone number they had owned throughout their history of mobile ownership and so their strategies would resort to ignoring phone calls and messages as a tactic to avoid harassment from loan-providing agencies and scammers. 

The younger participants approached this challenge by purchasing new alternate phone numbers, and then sought and used devices with dual SIM card capability, or handled multiple devices with separate use. While this will not erase their financial history, nor have an impact in improving/lessening their loan eligibility, it is a step that they found useful in addressing the harassment element.

\begin{quote}
    \textit{``I have separated my life. I use one line for calling and texting regularly, and use the other line for M-PESA and Equitel [bank] services. (P5)}
\end{quote}
The use of multiple phone lines had additional utility for avoiding overdraft repayment strategies. M-PESA has a \textit{fuliza} service affording users with the capability to overdraft their M-PESA account with a daily interest of 1\% and a repayment period of 30 days \cite{SafaricomFuliza}. After this period, any money received through M-PESA would automatically be used to defray the \textit{fuliza} balance before the recipient can access any remaining amount. 

\begin{quote}
    \textit{``I have two M-PESA accounts. The original one has fuliza [overdraft loan] on it. So I always have to be careful to specify which number I give to receive cash or airtime.'' (P4)} 
\end{quote}
The coping strategies that users adopt as a result of this landscape highlight the impact of trust not only to the specific offending services, but also on the whole financial landscape.

%============================================
%--------- SECTION: FINDINGS II -------------
%============================================

%\section{Impact on Agency and Privacy}
%---------------------------------------------
%**********************************************
\subsection{Impact on Agency and Privacy} 
%**********************************************
%**********************************************
%\subsection{Aging and Agency: Tension of Support} 
%**********************************************
Beyond the individualized trust signals and coping strategies, there are larger tension patterns that emerge: First, the notion of \textbf{agency}. Older participants were more likely to be targeted with, and believe SMS-type scams that request financial assistance, while the younger members were more likely to fall into loan/financial-access related scams. The eldest members leaned more on others to verify their transactions--distrusting their own judgment.

\begin{quote}
    \textit{``I stopped sending the money myself. I wait until [family] is around and have them help.'' (P17)}
\end{quote}
While this supported use provides a measure of trust, there is an emerging tension that this necessity erodes personal agency in addition to deepening dependency on others to understand nuances on scams. This upends the cultural community hierarchical order where elders are depended upon to dispense wisdom and guide actions--into elders viewed as dependents: stripping their dignity.

%**********************************************
%\subsection{Education and Access Tensions}
%**********************************************
Second, beyond the age considerations, the different participants exposure to the different types of fraud highlight the nuances between literacy (ability to read and write) and education (deductive reasoning and awareness). This is showcased by how smartphone users are still at a disadvantage to new exploitation mechanisms. 

\begin{quote}
    \textit{``You know, I am even suspicious of shopping apps and the like. A friend recommended this app but on install, it started sending messages to my contacts pretending to be me and asking for an amount of money. I mean, I didn't think that was even possible to be tricked like this and I have a university degree.'' (P3)}
\end{quote}
Only two participants knew how to search for the terms and conditions documents to one/multiple of the mobile-based application in their devices. Participants with smartphones had trouble parsing the language used in the terms and conditions documents. Participants with feature phones--who could read, could not access the document at all as their devices were not equipped with internet capability: highlighting the tension between \textbf{education} and \textbf{access}. 

%**********************************************
%\subsection{Technology and Data Suspicions}
%**********************************************
Third, the inaccessible language describing how the technical mechanisms of mobile-based transactions occur. This is in addition to the lack of transparency over how the systems work, information collected, and how that information is shared. This combination has served to increase suspicion and highlight the user burdens in managing the repercussions wrought by the shared information: chiefly, \textbf{technology and data suspicions}. 

\begin{quote}
    \textit{``They knew details about me, I don't know how they got them. I felt like I was being monitored and did not feel safe because I felt like I was very susceptible to lose my little cash.'' (P2)}
\end{quote}
The college student (P2) concerns surfaces not only the technological ability of the attackers, but also in the lack of clarity about data origins. People often infer their data to have originated from the app-based mobile service providers--given their rampant violation of privacy boundaries. The updated personal privacy law (2019) in Kenya is intended to protect against the unanticipated use of collected information. However, it fails to adequately address how personal information is collected/stored/reused, which we anticipate will have continued impact on the consumers' use of other measures of control.

%=============================================
%--------- SECTION:  DISCUSSION -------------- 
%=============================================
%To enfold in this discussion: 
%-- possible summarry table similar to what was presented in this paper?

 \section{Discussion} 
%---------------------------------------------

Through this work, we sought to understand how participants navigated the use of mobile-based financial infrastructure afforded by the M-PESA platform and service, how they established trust in operations, and handled the threats engendered by the close coupling of personal SIM card with finances. In this section, we situate our findings within other research: to critically examine the technology affordance, and use the insights to offer recommended steps to approach the scrutiny, the design and the responsibility for technology--especially those intended for developmental contexts. 

%**********************************************
\subsection{Interventions for Exploitation Disparities} 
%**********************************************

Access to financial instruments for the unbanked in Kenya has been supported by the easy availability of mobile-based micro loans. However, as the participant experience has shown, access to the financial infrastructure cannot be associated with equitable access. While mobile-based financial infrastructure addressed the scaling limitations inherent in the traditional microfinance infrastructure \cite{MutuaFaulu09} and table banking/\textit{chamas} \cite{KariukiCiteseer2014}, the profit motives have served to undermine aims inspired by both the UN Millennium and UN Sustainability goals. The prohibitively high interest rates \cite{FinancialStandard2018} compared with reasonable rates provided to those with access to conventional bank accounts \cite{CBK} highlight the key disparity: that the low-resourced often face unavoidable worse options.

Describing the mobile-based financial landscape amplified by the M-PESA platform underscored the policy aspects of the mobile-loan services and highlighted the fact that deliberately designed application actions--even when negatively impacting marginalized users, were in compliance with prevailing laws \cite{FinancialStandard2021, TechCrunch2021}. It is only when policies were amended and oversight instituted, that protective mechanisms were finally put into place. This action was only possible due to unceasing demands from stakeholders who were able to showcase systematic harms. We note this example to highlight that existing claimed ethical guidelines were used to exaggerate compliance and advantages of access, but did not explore the impact of harms, nor equitable access to financial infrastructure. Researchers have highlighted how traditional design decision-based signals such as app-permissions \cite{MunyendoSP2022} and terms of use documents while they have been unable to highlight the harms to the users, provide opportunities to audit and showcase systematic harms especially when the providers do not follow their documented policies \cite{Hindenburg2020}. The experience shared by participants further underscored their lack of access, lack of trust and lack of understanding of the terms and conditions documents: a feeling that extended to their felt lack of recourse in case of impacted finances. 

%-- where does design fit in this complex set of things we have talked about in this work: to be more sustainable, equitable and resilient. 

%**********************************************
%\subsection{Strategies for Risk Mitigation}  
%**********************************************

%--------- 
\subsection{Trust Signals and Strategies for Risk Mitigation} 
%--------------------------------------
Validation strategies leveraged by mobile users on \textbf{what} financial-related messages are legitimate include assigning trust to text messages originating from short-code numbers \cite{SafaricomCodes} and scrutinizing the serialization of receipt numbers on M-PESA receipts (as demonstrated in Figure \ref{fig:mpesa-scam}). These design and stylistic-based approaches are failure-prone to new deceptive strategies that thwart user scrutiny, or if the mobile provider amends how messages are conveyed. 

This landscape highlights the need for asking for caution on how new user requirements are effected. This is compounded by the tensions in policy updates intended to protect the users, the burdens that the compliance places on the most marginalized, and the opportunity for deceptions in the gap. For example, to protect user M-PESA account security that were risked by the use of counterfeit devices, the government in 2012 mandated that everyone had to own an IMEI\footnote{IMEI: International Mobile Equipment Identity-a unique numeric handset identifier.}-registered device, with a switch-off planned for unregistered devices \cite{ChebusiriBBC2012}. The most impacted were users who could not distinguish between counterfeit and authentic devices, and when those devices were switched off, they often could not afford new devices. Similar patterns emerged ten years later, when Kenyans were mandated to register their SIM cards in order to verify ownership and curb the use of SIM cards in fraudulent schemes \cite{TooStandard2022}. However, it was left to the service providers to enforce registration compliance. As a result, new verification deception tactics found success in targeting those without proper documentation or who could not physically travel to present their documentation to a service provider agent.

The participants experience underscore that the mechanisms of \textbf{how} the deceptive strategies work--are complicated by lack of sufficient explanation from official methods--often because fraudsters leverage lesser known USSD codes to lead users to unknowingly perform actions that make them liable for attacks such as SIM-swapping \cite{TongolaStandard2022}. The official guidance about these sophisticated attacks are vague \cite{TanuiStandard2022}, providing no specifics to aid the user in defending themselves \cite{Safaricom}. We found that even when the participants knew about the anatomy of a scam (i.e \textbf{what} it was), they could not articulate \textbf{how} it worked. What enforcement mechanisms exist are balkanized, because platforms are informally tasked with enforcing compliance. The prosecutorial powers are wielded only when large amount of money is lost. the severity engendering attention. There is an additional element of shame to being duped by these types of scams which serve to dampen the likelihood of victims reporting and proactively warning others. Participants often expect to absorb the losses--even when they use mechanisms provided by the platform to report scams for review and potential remediation. 

The issues brought about by the process of achieving policy compliance adds to the HCI research that has explored how people across different contexts have navigated the challenge. For example, highlighting the data concerns in the process of compliance whether this is expressed \cite{ShahICTD2019}, or reflected through user actions \cite{AhmedCHI2017}. This, in addition to the wrong assumptions and lack of uniformity in eliciting trust signals for determining trust of app-based systems \cite{MunyendoSP2022}, and the surfacing of lack of warning systems to users who face threats to their financial posture through text messages \cite{RazakCSCW2021}.

The three dimensions: (i) validation strategies to inform trust signals; (ii) how the users understood the deception to work; and (iii) their use of protective authorities to report cases of fraud, or to seek remedy; inform the user awareness and sensitivity to these types of financial threats and represent the goal for designers and researchers to address the policy gaps and support the users in ways to mitigate possible harms.

We found that successful risk mitigations revolved around verification, and involved users triangulating for trust--their methods dependent on the resources at hand: calling a second source, verifying with trusted circle, leveraging official dedicated service-provider lines, etc. This was true for both older adults who are likely to fall into deception involving threats towards policy compliance, and young adults who are more likely to fall victim as they seek access to income or mobile-based loans. The triangulation mechanisms should ideally operationalize steps from awareness campaigns, which in turn will demystify how the deceptive operations work. In this manner, design approaches will be accountable to users in supporting their agency, and augment their existing trust structures across literacy skills, education level and technology know-how. 

%-- where does design fit in this complex set of things we have talked about in this work: to be more sustainable, equitable and resilient. 

%**********************************************
\subsection{Supporting Harm Recovery} %Designing for Transparency (RQ 3)
%**********************************************

HCI research has highlighted how the use of off-the-shelf technology is a poor fit for use in developmental contexts \cite{BrewerComputer2005}. The building of M-PESA showcased a successful case that was built based on learning from how people locally approach financial transactions \cite{HughesMIT2007}. Its sustained success is a testament for this design approach. However, the acceleration of the financial tech space, and the profit motives submerging developmental goals have served to undermine any original design intent--and often disadvantages users long after they have ceased using the technology. This is observed by how these approaches have spurred changes in user behaviors to protect their privacy and agency: in some cases, leading them to view technology with suspicions or abandoning use altogether.

Participants' needs on supporting harm recovery involve two policy-driven aspects: restoring lost funds to users; and in penalizing the abusers. The latter is now being effected in Kenya: the new policies guiding charges and fines for breaching data laws \cite{MwangiBD2023} and on deceptive scams perpetuated by individuals \cite{TanuiStandard2022}. They form the genesis of justice process, but there is a long way to go to identify the marginalized users and enact fair adjudications. There is need curb the shame in this domain and processes: by demystifying how scams operate, illustrating the underlying schemes, creating awareness of deceptive patterns, showcasing scale, and creating redundancies in support by empowering users with knowledge--strategies that all fall within designers' purview. This can be done in addition to reevaluating how to explain terms to users with no access to internet, or an understanding of the English language--and ultimately exposing when platforms act against standing policies and the terms and conditions they provide to users. 

%**********************************************
\subsection{Other Design Guidelines and Implications}
%**********************************************

Researchers have identified mechanisms for conducting research with underrepresented communities, and have studied the impact of technology design for/with these communities from postcolonial \cite{IraniCHI2010} and decolonial approaches \cite{SadiAfriCHI2021}.  For design mechanisms that offer services and/or audits to be successful, the metrics of success need to transcend simply ``poverty reduction'' and to consider users' mental models regardless of their literacy level: a measure of awareness; of understanding of the threat landscape; and importantly, how they can defend themselves--as means of supporting their agency in decision making and therefore their own pathways for addressing poverty.  

For design-based recommendations to be effective, they have to be undergirded by on-the-ground policies. However, as we have found, the policies are often enacted when there is sufficient proof or a crisis point. In this way, design approaches should be in service to the users by offering knowledge, support and infrastructure: demystifying the decision-making structures, easing the process of understanding compliance, and providing comparisons--highlighting existing disparities. This meta approach we argue, beyond providing reference and tools for users to advocate for themselves, would also be in service of policy makers to determine courses of action that do not depend on sufficient marginalized users being harmed as an impetus and proof. 

Successful and reparatory design-based recommendations ought to lean into the supportive financial infrastructure and organization: learning from similar approaches leveraged in locally-organized table-banking/\textit{chamas} and how the group sustains and supports individual members. This should be in addition to the spirit of microfinance offerings--with the intent to empower the borrowers context-specific financial advice as well as providing access to capital at fair rates.

The affordance tensions presented in this paper represents ongoing struggles with development and advantages of access, and highlight the cyclical patterns of how harms are effected. To wit, initial projects are provided with development agenda, then the profit opportunity become apparent as new users are identified, after which new providers emerge to take advantage of the market, followed by the profit motives causing harm to same users who were marginalized. Therefore, new/updated policies are effected to mitigate harms... then the cycle repeats. The case study on mobile-based loans showcases how this issue cycle has abided since 2013--and can be considered as a clarion call to examine other dimensions of provisions intended for marginalized users \cite{WatuCredit} and the emerging concerns about disparities enabled through design. 

%**********************************************
\subsection{Limitations and Future Work}
%**********************************************

We specifically recruited participants with experience of tech-enabled harms in the context of the Kenyan financial landscape, and further restricted this to those who had experience with borrowing money through M-PESA, financial applications that scaffold on the M-PESA service, or applications relying on traditional banking service. We included a diversity of users across urban and rural areas, with various literacy and socio-economic backgrounds and technological know-how. These inclusion/exclusion criteria resulted in 17 participants. While this was useful for coherence and depth in understanding of the financial harms, the sample size is limited, and may impact generalizability across other financial experiences. 

Future work should consider the broader financial landscape, increasing the possibility to provide nuances of the financial offerings, and providing comparisons across individual, group, and systematic-based contexts. This focus will provide an understanding of the financial landscape and asymmetric impacts, and make possible longitudinal analysis and sufficient scale for comparisons with other financial landscapes outside of the Kenyan context.

%=============================================
%--------- SECTION:  CONCLUSION--------------
%=============================================

\section{Conclusion}
%---------------------------------------------

We explored the usage of mobile-based money transfer and loan servicing applications: the affordances and access it provides to the marginalized, but also the weakness of policy and ethical guardrails to protect users on account of fraud and scams. We interviewed 17 participants who use these services to understand their contexts of use, the challenges that they face, and to highlight the gaps in policy, research and design presenting insights on individual actions based on evasion of harm, harm encounter, and harm recovery.

We presented strategies for contextual design considerations learned from how participants use their devices. This was borne out of how they interact with prevailing systems--even as they are aware that they are being harmed in the process--but having no choice in the matter. This highlights the dearth of governing policies to support users, and provide opportunities for researchers and designers to offer utilities in support of user awareness, and to aid policy-makers by creating tools and designs to offer clarity in the performance of existing protective guardrails, and where needed, spotlight the necessity for new approaches to address evolving technology.

%==========================================================================================
%=================================== END OF PAPER =========================================
%==========================================================================================

\bibliographystyle{ACM-Reference-Format}
\bibliography{paper-ref}

\end{document}